\documentclass[10pt, conference, compsocconf]{IEEEtran}
\ifCLASSINFOpdf
\else
\fi
\usepackage{booktabs}
\usepackage{amsmath}
\usepackage{amssymb}
\usepackage{bm}
\usepackage{etoolbox}
\usepackage{array,ragged2e}
\usepackage{tikz}
\usepackage{csquotes}
\usetikzlibrary{decorations.pathreplacing}
\usetikzlibrary{automata,positioning}
\usepackage{graphicx}
\graphicspath{{./}}
\usepackage[none]{hyphenat}
\usepackage{array}
\usepackage{xspace}
\newcommand{\etal}{\emph{et al.}\xspace}
\usepackage[backend=biber,style= numeric-comp,sorting=none, maxbibnames=99]{biblatex}
\renewbibmacro{in:}{%
	\ifentrytype{article}{}{\printtext{\bibstring{in}\intitlepunct}}}
\usepackage{environ}
\usepackage{caption}

\DeclareMathOperator*{\E}{\mathbb{E}}

\begin{document}
\title{Analysing The Impact Of A DDoS Attack Announcement On Victim Stock Prices}
\date{}

\author{\IEEEauthorblockN{Abhishta}
	\IEEEauthorblockA{University of Twente\\
		Enschede, The Netherlands\\
		Email: s.abhishta@utwente.nl}
	\and
	\IEEEauthorblockN{Reinoud Joosten}
	\IEEEauthorblockA{University of Twente\\
		Enschede, The Netherlands\\
		Email: r.a.m.g.joosten@utwente.nl}
	\and
	\IEEEauthorblockN{L.J.M. Nieuwenhuis}
	\IEEEauthorblockA{University of Twente\\
		Enschede, The Netherlands\\
		Email: l.j.m.nieuwenhuis@utwente.nl}}


%Telephone: (800) 555--1212, Fax: (888) 555--1212}



	\maketitle
	
	\begin{abstract}
		DDoS attacks are increasingly used by `hackers' and `hacktivists' for various purposes. A number of on-line tools are available to launch an attack of significant intensity. These attacks lead to a variety of losses at the victim's end. We analyse the impact of Distributed Denial-of-Service (DDoS) attack announcements over a period of 5 years on the stock prices of the victim firms. We propose a method for event studies that does not assume the cumulative abnormal returns to be normally distributed, instead we use the empirical distribution for testing purposes. In most cases we find no significant impact on the stock returns but in cases where a DDoS attack creates an interruption in the services provided to the customer, we find a significant negative impact.
	\end{abstract}
	
	
	\section{Introduction}
	The trend of significant growth in the magnitude of high intensity DDoS attacks has been consistent in the past years \cite{WISR2015} and these attacks have resulted in heavy losses for firms \cite{PI2015}. The rise in the number of attacks being encountered can be attributed to the ample availability of online tools for launching DDoS attacks. Booter websites have become successful in creating a market for themselves and as a consequence technical knowledge is no longer a prerequisite for launching a DDoS attack \cite{Santanna2014}.
	
	The losses encountered by firms due to these cyber assaults can be divided into direct and indirect ones \cite{Anderson2012}. Financial damages due to infrastructural downtime, loss of online traffic, paid ransom and customer compensation etc. are accounted as direct losses. Indirect losses include damage to company's reputation and impact at stock prices etc. We examine the indirect loss due to the decrease in the market value of a firm as a result of an announcement of getting hit by a DDoS attack. In Section \ref{previous works1} we discuss several studies on the impact of information security breaches on the stock prices \cite{Spanos2016}.
	
	In our study, we consider all DDoS attacks reported after 2010. We do this in order to understand the effects caused by these announcements. Unlike earlier studies we will study the impact of DDoS attack announcements only, because these attacks do not lead to any form of information leaks and do not pose any danger to customer data. Hence, in our sample we do not consider any of the events where DDoS has been used as a smoke screen.\footnotetext[1]{This paper has been published in the proceedings of 2017, 25th Euromicro International conference on Parallel, Distributed, and Network-Based Processing (PDP). This is a pre-final version, publisher made some changes before publication.}
	
	\section{Previous Work}
	\label{previous works1}
	
	\begin{center}
		\begin{table*}[!ht]
			\caption{Previous works on impact on victim stock prices.
				\label{Previous Works}}
			\begin{tabular}{c c c c c >{\centering}m{4.4 cm} c}
				\midrule
				& \textbf{Author} & \textbf{Estimation Model} & \textbf{Sample Size} & \textbf{Breach Type} & \textbf{Conclusion} & \textbf{Sample Period}\\
				\midrule
				\cite{HovavAnatandDArcy2003} & Hovav \& D'Arcy (2003) & Market Model & 23 & DoS & No significant impact of DoS attacks on the capital market. Some indication of impact on firms that rely on the web for their business. & 1998-2002\\
				\midrule
				\cite{Campbell2003} & Campbell \etal (2003)  & Market Model & 43 & Generic & Some negative stock market impact to reported information security breaches. & 1995-2000\\
				\midrule
				\cite{Garg2003} & Garg \etal (2003)  & N/A & 22 & Generic & Average fall in the stock price was approximately 2.9\% over a 2-day and 3.6\% over 3-day period. & 1996-2002\\
				\midrule
				\cite{Cavusoglu2004} & Cavusoglu \etal (2004)  & Market Model & 66 & Generic & Security breach announcements affect the values of the announcing firms and also the Internet security developers. & 1996-2001\\
				\midrule
				\cite{Kannan2007} & Kannan \etal (2007)  & Market Model & 102 & Generic & Drop of 1.4\% in the market valuation relative to the control group of firms. & 1997-2003\\
				\midrule
				\cite{Gordon2011} & Gordon \etal (2011) & Fama-French Model & 258 & Generic & Pre 9/11 information security breaches showed significant negative stock market returns but the results for the post 9/11 period were not significant. & 1995-2007\\
				\midrule
			\end{tabular}
		\end{table*}
	\end{center}
	
	Event studies measure the impact of company related events on the market value of the firm. MacKinlay \cite{Mackinlay1997} discusses the procedure for conducting an event study and also the various models that can be used for estimation of normal behaviour of the market. In the past many researchers have studied the impact of information technology related events on the market value of the firm. Santos \etal \cite{Santos1993} examined the impact of information technology investment announcements on the market value of the firm and suggested that there is no significant impact of these investment announcements on the market value.
	
	Previous studies \cite{HovavAnatandDArcy2003,Campbell2003,Cavusoglu2004,Kannan2007} have used a one-factor market model for the estimation of stock prices as shown in Equation \ref{market model}. Where $r_{it}$ represents the rate of return of the stock $i$ and $r_{mt}$ represents the rate of return of the market index on day $t$. For instance, $r_{it}$ can be calculated as $(P_{it}-P_{it-1})/P_{it-1}$, where $P_{it}$ is the price of the stock on day $t$. 
	
	\begin{equation}
		\label{market model}
		r_{it}={\alpha_i}+{\beta_i}r_{mt} + \epsilon_{it}
	\end{equation} 
	
	The parameters $\alpha$ and $\beta$ are firm dependent coefficients. $\hat{\alpha}$ and $\hat{\beta}$ are their ordinary least square (OLS) estimators. The stochastic variable $\epsilon_{it}$ is the error term with $\E{[\epsilon_{it}]}=0$. Gordon \etal \cite{Gordon2011} uses a Fama-French three factor model \cite{famafrench} to predict the stock prices. The three factors being company size, company price-to-book ratio and market risk. The three factor model is shown in Figure \ref{famafrench}. 
	
	\begin{equation}
		\label{famafrench}
		r_{it}={a_i}+{b_i}r_{mt}+{s_i}SMB_t+{h_i}HML_t+\epsilon_{it},	
	\end{equation}
	
	$SMB_t$ is the difference between the return on the portfolio of small stocks and the return on the portfolio of large stocks on day $t$ and $HML_t$  is the difference between the return on a portfolio of low-book-to-market stocks and the return on a portfolio of low-book-to-market stocks on day $t$. The parameters ${a_i}$,${b_i}$,${s_i}$ and ${h_i}$ are Fama and French three-factor model estimated firm dependent coefficients. The stochastic variable $\epsilon_{it}$ is the error term with $\E{[\epsilon_{it}]}=0$.
	
	These studies \cite{Kannan2007,HovavAnatandDArcy2003,Campbell2003} use abnormal returns (additive) and cumulative abnormal returns (additive) as a measure of event impact. Equations \ref{addar} and \ref{addcar} show the relations used to compute abnormal returns and cumulative abnormal returns respectively. As they assume normal distribution for the $CAR$ values hence they use Z statistic to test their hypothesis.
	
	\begin{equation}
		\label{addar}
		AR_{it}=r_{it}-(\hat{\alpha_i}+\hat{\beta_i}r_{mt})
	\end{equation}
	
	\begin{equation}
		\label{addcar}
		CAR_n=\sum_{t=-1}^{n}AR_{it}
	\end{equation}
	
	Past studies have been conducted on evaluating the impact of information security breaches on the prices of the victim firm's shares. Table \ref{Previous Works} lists selected works and their conclusions. In this table we also take a look on the sample size and period of the sample considered by these studies.  
	
	Previous studies had a mixed response on the impact of denial of service attacks on the stock returns of the victim firms. Some studies like Garg \etal \cite{Garg2003} and Hovav \& D'Arcy \cite{HovavAnatandDArcy2003} suggest that DDoS attack announcements lead to a negative abnormal returns, while Gordon \etal \cite{Gordon2011} deny the effect of these attacks on the market value of the firm. Spanos \& Angelis \cite{Spanos2016} conducted a systematic literature review on the impact of information security events on the stock market and concluded that the events examined created a significant impact on the stock price of the firms.

	\section{Method}
	
	\begin{figure}[!ht]
		\begin{center}
			\begin{tikzpicture}[scale=0.8,shorten >=1pt,node distance=5cm,on grid,auto]
			\node[text width=3cm,align=center] (data) at (-2,0)  {Historical stock \\ Prices($R_i$) \\ (200 days)};
			\node[text width=3cm,align=center] (data1) at (2,0)  {S\&P 500 \\ Index Values ($R_m$)};
			\node[text width=3cm,align=center] (rate) at (0,-2) {Calculation of rates ($r_i,r_m$)};
			\node[text width=4cm,align=center] (market) at (0,-4) {\emph{Multiplicative model}};
			\node[text width=4cm,align=center] (abnormal) at (0,-6) {Calculation of abnormal returns $AR_i$};
			\node[text width=4cm,align=center] (scenario) at (0,-8) {\emph{Generation of random scenarios}};
			\node[text width=4cm,align=center] (actual scenario) at (0,-10) {Determining the position of actual scenario};
			\draw [->] (data)->(rate);
			\draw [->] (data1)--(rate);
			\draw [->](rate)--(market);
			\draw [->](market)--(abnormal);
			\draw [->](abnormal)--(scenario);
			\draw [->](scenario)--(actual scenario);
			\draw [decorate,decoration={brace,amplitude=10pt},xshift=-2pt,yshift=0pt](-3.5,-1) -- (-3.5,0.75) node [black,midway,xshift=-0.6cm]{\footnotesize {Data Collection}};
			\draw [decorate,decoration={brace,amplitude=10pt},xshift=-2pt,yshift=0pt](-3.5,-10) -- (-3.5,-1) node [black,midway,xshift=-0.6cm]{\footnotesize {Analysis}};
			
			\end{tikzpicture}
		\end{center}
		\caption{Method for event study. (Our contribution in \emph{Italics}.)}
		\label{Method Diagram}
	\end{figure}
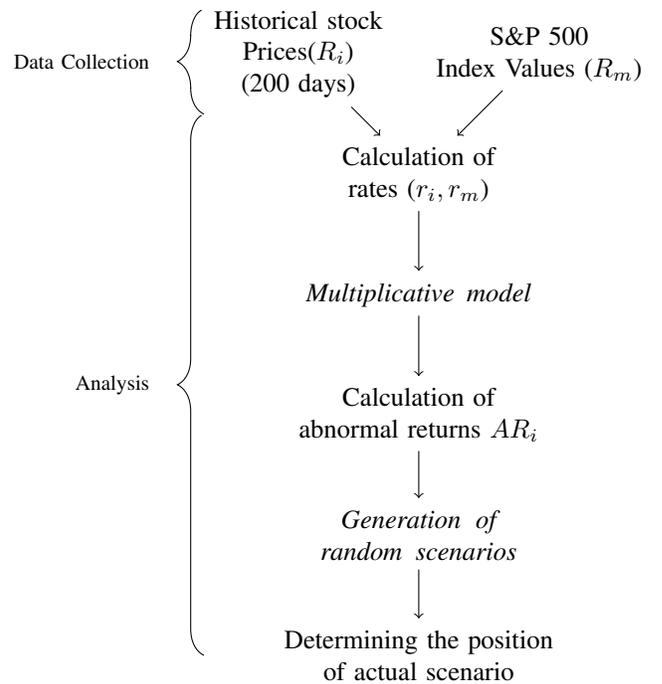  
	
	To analyse the impact of DDoS attack announcements on stock returns we use the method as shown in Figure \ref{Method Diagram}. We can broadly divide the method into two sections:
	
	\begin{enumerate}
		\item Data collection.
		\item Analysis
	\end{enumerate}
	
	Our contribution to the analysis is at two instances. Firstly, we use a \emph{multiplicative model} for the estimation of return rates and secondly, we use the empirical distribution of abnormal returns by \emph{generation of random scenarios} for the analysis. In Section \ref{Sec: Data Collection} we explain the approach for data collection. Section \ref{Sec: Analysis} deals with the identification of the impact caused by the announcements.

	\subsection{Data Collection}
	\label{Sec: Data Collection}
	
	\begin{table*}[!htbp]
		\caption{Sample of DDoS attack events.}
		\begin{center}
			\begin{tabular}{c c c c c}
				\midrule
				\textbf{Organisation} & \textbf{Announcement Date} & \textbf{Source} & \textbf{Infrastructure} & \textbf{Firm Type}\\ 
				\midrule
				Master Card & 8-12-2010 & spiegel.de & Website & Financial Services\\
				Visa & 8-12-2010 & spiegel.de & Website & Financial Services\\
				Bank of America & 27-12-2010 & infosecisland.com & Website & Financial Services\\
				Vodafone & 5-10-2011 & infosecurity-magazine.com & None & Telecommunications\\
				Apple & 29-5-2012 & att-iphone-unlock.com & Website & IT\\
				AT\&T & 16-8-2012 & pcworld.com & None & Telecommunications\\
				Wells Fargo & 20-12-2012 & technologybanker.com & DNS & Financial Services\\
				JP Morgan Chase & 13-3-2013 & scmagazine.com & Website & Financial Services\\
				TD Canada Trust & 21-3-2013 & thestar.com & E Services & Financial Services\\
				American Express Company & 28-3-2013 & bankinfosecurity.com & Website & Financial Services\\
				International Netherlands Group & 9-4-2013 & nrc.nl & Payment Services & Financial Services\\
				Linkdin & 21-6-2013 & news.softpedia.com & Website & Social Networking\\
				Microsoft & 27-11-2013 & scmagazine.com & DNS & IT/Gaming\\
				Royal Bank of Scotland & 4-12-2013 & theguardian.com & Banking Services & Financial Services\\
				JP Morgan Chase & 30-1-2014 & bobsguide.com & Online Banking Services & Financial Services\\
				Bank of America & 30-1-2014 & bobsguide.com & Online Banking Services & Financial Services\\
				Facebook & 21-2-2014 & nos.nl & Messageing Services & Social Networking\\
				Activision Blizzard & 29-3-2014 & ign.com & Gaming Services & Gaming\\
				Danske Bank & 10-7-2014 & ddosattacks.net & Website & Financial Services\\
				Storebrand & 10-7-2014 & ddosattacks.net & Website & Insurance Company\\
				Gjensidige Forsikr & 10-7-2014 & ddosattacks.net & Website & Insurance Company\\
				Sony Corporation & 24-8-2014 & techcrunch.com & Gaming Services & IT\\
				Amazon & 27-8-2014 & shacknews.com & Twitch Streamers & E-commerce\\
				Activision Blizzard & 14-11-2014 & eurogamer.net & Gaming Services & Gaming\\
				Sony Corporation & 26-11-2014 & wiwo.de & Gaming Services & IT\\
				Rackspace & 22-12-2014 & welivesecurity.com & DNS & Hosting\\
				Microsoft & 24-12-2014 & krebsonsecurity.com & Gaming Services & IT/Gaming\\
				Sony Corporation & 24-12-2014 & krebsonsecurity.com & Gaming Services & IT\\
				Alibaba Group & 25-12-2014 & ddosattacks.net & Cloud Services & E-commerce\\
				Nordea Bank & 4-1-2015 & ddosattacks.net & Online Banking Services & Financial Services\\
				Facebook & 27-1-2015 & gizmodo.com.au & Website & Social Networking\\
				Amazon & 16-3-2015 & scmagazineuk.com & Twitch Streamers & E-commerce\\
				EA Sports & 18-3-2015 & ibtimes.com & Gaming Services & Gaming\\
				Ziggo & 18-8-2015 & emerce.nl & DNS & Telecommunications\\
				Overstock.com & 3-9-2015 & ddosattacks.net & DNS & E-commerce\\
				\hline
			\end{tabular}
		\end{center}
		\label{sample}
	\end{table*}
	
	In this study we consider all DDoS attack announcements that were made on the web since `Operation Payback', launched by Anonymous in December, 2010. Table \ref{sample} shows the final list of all announcements that we analysed. For each attack we collected the date of announcement, the company type and also the services disrupted. The initial list consisted of 43 announcements.
	
	We further filtered the list using the following criteria:
	
	\begin{enumerate}
		\item If multiple announcements were made on consecutive days, then the earliest date was considered.
		\item All announcements regarding companies that were not publicly traded at the time of attack were eliminated.
		\item All attack announcements in which DDoS attack was coupled with information theft were also not considered for analysis. This was done in order to be able to see the isolated effect of a DDoS attack announcements on the firm's stock price.    
	\end{enumerate}
	
	The stock prices for all the firms in the sample were collected by using the Yahoo! finance API. For measuring the market rate we collected the S\&P 500 index values.The final sample consisted of 35 announcements.

	\subsection{Analysis}
	\label{Sec: Analysis}
	
	We depart from the familiar research strategy for event studies for the following reasons. We wish to avoid the widespread practice of approximating multi-day returns by simply adding up the corresponding single-day returns\footnote{Note that a 10\% increase, followed by a 10\% decrease imply a total decrease of 1\% according to the multiplicative formula $(1.1)(0.9)=0.99$. The additive approximation would yield a 0\% change, an overestimation of 1\%.} and instead use the exact ones. Secondly, we want to avoid the equally wide-spread assumption about short-term returns being (approximately) distributed according to a normal, i.e., Gaussian, distribution. We refrain from imposing as an alternative one of the better known distributions such as the Weibull or the Erlang distributions, as the problem generally is not only skewness (asymmetry) but fatness of both tails, i.e., realisations quite far from the average are more common than for instance in the normal distribution with the same mean and variance. Another route not taken is to use the data consisting of a sample of returns for a period of 200 days prior to the event, and tone of these alternative distributions to the data. Instead we assume that the one-day returns follow an unknown distribution which we are going to approximate by the empirical distribution, i.e., the distribution of the 200-day-sample returns.
	
	We acknowledge the considerable merits of the widespread research strategy involving these two approximations as they subsequently allow the construction of test variables which follow the Student's t-distribution in order to engage in the testing of hypotheses. 
	
	As we do not use the approximations central to this research strategy, we are faced with the challenge of establishing a pertinent distribution for similar hypothesis testing. We do this by the technique of bootstrapping (e.g., Efron \cite{efron1992bootstrap}) which in our case involves generating a sufficiently large number of multi-day returns by drawing from the empirical distribution a number of consecutive one-day returns corresponding to the number desired. With such a series of one-day returns we compute the exact multi-day returns, and proceed in the same fashion to obtain a large number (in our case 5 million) of such multi-day returns. The relative frequencies of this large population of exact multi-day returns are then employed as the relevant distribution for hypothesis testing. Note that the standard approach in event studies is to take as the null hypothesis that the event has no influence at all, meaning in statistical terms that the distribution of returns before the event and the one after the event are identical. So, under this assumption the sample returns can be indeed used to generate the relevant distributions of multi-day returns.
	
	We consider a multiplicative model to represent the normal behaviour of the market. According to the model if $r_{it}$ represents the rate of return of the stock $i$ on day $t$ and $r_{mt}$ represents the rate of return of the market index on day $t$, then the model can be represented mathematically by Equation \ref{model}. Rate of return can be calculated as shown in Equation \ref{rate}, where $R_{it}$ and $R_{mt}$ represent the stock price and market index for day $t$. The value of the market index shows the average of returns of all the firms included in the market index.

	\begin{equation}
		\label{rate}
		\begin{split}
			r_{it} &= \dfrac{R_{it}-R_{i(t-1)}}{R_{i(t-1)}}\\
			r_{mt} &= \dfrac{R_{mt}-R_{m(t-1)}}{R_{m(t-1)}}\\
		\end{split}
	\end{equation}
	
	\begin{equation}
		\label{model}
		(1+r_{it}) = \alpha_i(1+r_{mt})^{\beta_i}\\
	\end{equation}

	A multiplicative model\footnote[1]{If the returns $r_{it}$ and $r_{mt}$ are extremely small a linear relationship between these variable will give a good approximation.} is used to estimate the returns on a firm's stock. In this study we use the Standard and Poor's (S\&P) 500 as the index of the market. S\&P 500 has been used as a market study in many of the previous event studies. The parameters $\alpha_i$ and $\beta_i$ are firm dependent and will be estimated.
	
	Equation \ref{model} is linearised in Equation \ref{log model}. The stochastic variable $\epsilon_{it}$ is the error term with $\E{[\epsilon_{it}]}=0$. We use ordinary least square (OLS) estimation to obtain estimations $\widehat{\ln{\alpha_i}}$ and $\hat{\beta_i}$ for $\ln{\alpha_i}$ and $\beta_i$ by considering daily returns over a period of 200 days. This period starts 201 days before the date of announcement and ends 2 days before the announcement. In this study we will call this the \emph{estimation period}. This length of the estimation period is consistent with the previous event studies \cite{Gordon2011,HovavAnatandDArcy2003,Santos1993}.
	
	\begin{equation}
		\label{log model}
		\ln(1+r_{it}) = \ln(\alpha_i)+\beta_i\ln(1+r_{mt})+\epsilon_{it}
	\end{equation}
	
	The abnormal return measures the deviation that the stock shows from the model we calculate. $AR$ is calculated for the estimation period and is given by Equation \ref{abnormal returns}. Hence, abnormal returns can be calculated by using Equation \ref{transformed}.

	\begin{equation}	
		\label{abnormal returns}
		\ln(1+AR_{it})=[\ln(1+r_{it})-[\widehat{\ln(\alpha_i)}+\hat{\beta_i}\ln(1+r_{mt})]]\\
	\end{equation}
	
	\begin{equation}
		\label{transformed}
		AR_{it} = \frac{(1+r_{it})}{\hat{\alpha_i}(1+r_{mt})^{\hat{\beta_i}}}-1\\
	\end{equation}
	
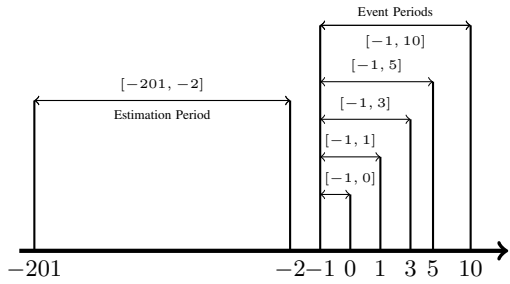
\begin{figure}[t]
		\begin{center}
			\begin{tikzpicture}
			\draw[->,ultra thick] (-5,0)--(1.5,0);
			\draw[-,thick] (-4.8,0)node[below]{\small{$-201$}}--(-4.8,2) ;
			\draw[-,thick] (-1,0)node[below]{\small{$-1$}}--(-1,3) ;
			\draw[-,thick] (-1.40,0)node[below]{\small{$-2$}}--(-1.40,2) ;
			\draw[-,thick] (0.5,0)node[below]{\small{$5$}}--(0.5,2.25) ;
			\draw[-,thick] (1,0)node[below]{\small{$10$}}--(1,3) ;
			\draw[-,thick] (-0.6,0)node[below]{\small{$0$}}--(-0.6,0.75);
			\draw[-,thick] (-0.2,0)node[below]{\small{$1$}}--(-0.2,1.25);
			\draw[-,thick] (0.2,0)node[below]{\small{$3$}}--(0.2,1.75);
			\draw[<->] (-4.8,2)--(-1.40,2) node[midway,below]{\tiny{Estimation Period}};
			\draw[<->] (-4.8,2)--(-1.40,2) node[midway,above]{\tiny{$[-201,-2]$}};
			\draw[<->] (-1,3)--(1,3) node[midway,below]{\tiny{$[-1,10]$}};
			\draw[<->] (-1,2.25)--(0.5,2.25) node[midway,above]{\tiny{$[-1,5]$}};
			\draw[<->] (-1,0.75)--(-0.6,0.75) node[right,above]{\tiny{$[-1,0]$}};
			\draw[<->] (-1,1.25)--(-0.2,1.25) node[midway,above]{\tiny{$[-1,1]$}};
			\draw[<->] (-1,1.75)--(0.2,1.75) node[midway,above]{\tiny{$[-1,3]$}};
			\draw[<->] (-1,3)--(1,3) node[midway,above]{\tiny{Event Periods}};
			\end{tikzpicture}
		\end{center}
		\caption{Estimation and Event Periods.}
		\label{periods}
	\end{figure}

	The estimator\footnote{Note that $\hat{\alpha}$ is not $e^{\widehat{\ln(\alpha)}}$ as $\E{[\alpha]}\neq\E{[\ln{\alpha}]}$.} $\widehat{\ln(\alpha)}$ is good for $\ln(\alpha)$ but cannot be used to estimate $\hat{\alpha}$. Hence, for estimating $\hat{\alpha}$ we make use of Equation \ref{alpha estimator}, that is derived using Equation \ref{log model}.
	
	\begin{equation}
		\label{alpha estimator}
		\hat{\alpha_i}=\dfrac{\sum_{t=1}^{T}(1+r_{it})}{\sum_{t=1}^{T}(1+r_{mt})^{\hat{\beta_i}}},
	\end{equation}
	
	\noindent where $T$ is the total number of days in the estimation period. In order to measure the impact of the announcements on the stock return we define five event periods as shown in Figure \ref{periods}. These are:
	
	\begin{enumerate}
		\item One day prior to the announcement to the day of announcement $[t-1,t]$.
		\item One day prior to the announcement to 1 days after it $[t-1,t+1]$.
		\item One day prior to the announcement to 3 days after it $[t-1,t+3]$.
		\item One day prior to the announcement to 5 days after it $[t-1,t+5]$.
		\item One day prior to the announcement to 10 days after it $[t-1,t+10]$.
	\end{enumerate}
	
	We consider the starting point of the event period one day prior to the announcement so as to accommodate for information leaks. We randomly draw 2,3,5,7 and 12 abnormal returns from the estimation period to represent the abnormal returns for event periods $[-1,0],[-1,1],[-1,3],[-1,5]$ and $[-1,10]$ respectively. We generate five million possible random scenarios for the event periods. Recall, we do not assume the abnormal returns to be normally distributed at any point. This is done to improve the accuracy of our results. We consider short event periods of 2 days, 3 days, 5 days, 7 days and 12 days respectively in accordance with the results of previous studies \cite{Garg2003,HovavAnatandDArcy2003,Gordon2011}. 
	
	For evaluating the combined effect over a certain number of days we also calculate cumulative abnormal returns for the randomly generated scenarios. $CAR$ is calculated using relation shown in Equation \ref{CAR}. Where, $N_1$ and $N_2$ represent the start and ending days of the event period. The actual $AR$s and $CAR$s for the event period are calculated using Equations \ref{abnormal returns} and  \ref{CAR} respectively on the real stock data for the event periods. It is important to note that previous studies have assumed these cumulative abnormal returns to be normally distributed for strategic convenience. We use the empirical distribution of $CAR$ for analytical purposes, i.e. for hypothesis testing.
	
	\begin{equation}
		\label{CAR}
		CAR= \prod_{t=N_1}^{N_2}(1+AR_{it})-1
	\end{equation}
	
	\begin{figure}[t]
		\centering
		\includegraphics[width=0.5\textwidth]{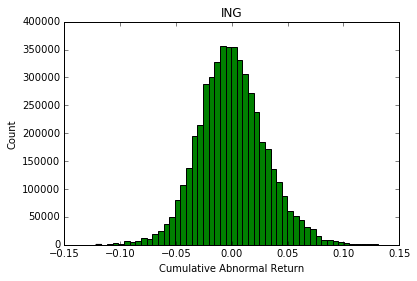}
		\caption{Empirical distribution of two-day $CAR$ values for ING.}
		\label{fig:distribution}
	\end{figure}

\begin{figure}[h!]
		\begin{center}
			\begin{tikzpicture}[shorten >=1pt,node distance=5cm,on grid,auto]
			\node[text width=3cm,align=center] (CAR) at (-2.7,0)  {$CAR$};
			\node[text width=5cm,align=center] (N) at (3,0)  {No Significant Impact};
			\node[text width=5cm,align=center] (P) at (3,1)  {Positive Impact};
			\node[text width=5cm,align=center] (Ne) at (3,-1)  {Negative Impact};
			\draw (-1,1)--(-1,0)[-];
			\draw [-] (-1,0)--(-1,-1);
			\draw [-] (-2,0)--(-1,0);
			\draw [->] (-1,0)--(1,0)node[midway,above]{\tiny{$10<\%ile<90$}};
			\draw [->] (-1,-1)--(1,-1)node[midway,above]{\tiny{$CAR<0$}};
			\draw [->] (-1,-1)--(1,-1)node[midway,below]{\tiny{$\%ile<10$}};
			\draw [->] (-1,1)--(1,1)node[midway,above]{\tiny{$CAR>0$}};
			\draw [->] (-1,1)--(1,1)node[midway,below]{\tiny{$\%ile>90$}};
			\end{tikzpicture}
		\end{center}
		\caption{Decision Rule.}
		\label{Rule}
	\end{figure}
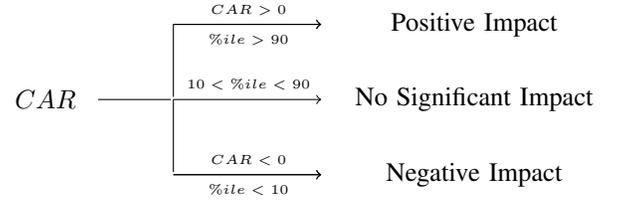

	Finally, to determine the effect of the announcement on the daily stock return rates we check where do the cumulative actual abnormal returns lie in the distribution of simulated cumulative abnormal returns (multiplicative). Figure \ref{fig:distribution} shows an example of the distribution two-day $CAR$ values for \emph{International Netherlands Group}. A highly unlikely negative $CAR$ value will represent a negative impact of the announcement and the actual scenario will fit to the extreme left of the probability distribution. For our analysis we consider 10 percentile of the scenarios on the left to be representative of a negative impact and 10 percentile of the scenarios on the right represent the positive impact. Hence, for the evaluation of the results we use the decision rule as shown in Figure \ref{Rule}.
	
	\section{Results}
	
	The complete results for our study are shown in Table \ref{Full Results} in Appendix \ref{Results Table}. According to the results of our analysis we observe a significant negative impact in the case of \emph{International Netherlands Group} and \emph{EA sports}. Whereas, a delayed negative effect is noticeable in the case of \emph{Bank of America}, \emph{Storebrand} and \emph{Nordea Bank}. In most cases we do not see a negative effect on the victim stock prices. 
	
	In cases where the announcements state that the availability of the infrastructure under attack did not affect the customers, no significant impact was noticed. For example, in the case of \emph{Visa} and \emph{MasterCard} the infrastructure under attack was their \emph{website} but the customers were still able to use their cards for payment purposes. Whereas in the case of \emph{International Netherlands Group}, customers had troubles using the payment services. Similarly, in the case of \emph{EA Sports}, gamers were not able to log onto their on-line gaming accounts.
	
	In the case of \emph{Ziggo}, the customers did face troubles due to the unavailability of internet services but as the firm is a part of a bigger conglomerate \emph{Liberty Global}, we were unable to spot any significant impact on the stock prices. 
	
	\section{Conclusion} 
	
	As a conclusion, we can say that there is a noticeable negative impact on the stock prices of the victim firm whenever the attack causes interruptions to the services provided by the firm to its customers. This drop is consistent with the results of the previous studies \cite{Garg2003,HovavAnatandDArcy2003}. However, it is not possible to comment on the intensity of the impact because it is firm dependent.
	
	\printbibliography
	
	\clearpage

	\onecolumn
	\centering
	\appendix{\emph{A: Complete Results.}}
	\section{Complete Results.}
	\label{Results Table}
	\centering
	%\large\textbf{Complete Results.}
	
	\begin{table}[h]
		\centering
		\caption{Results of analysis.}
		\label{Full Results}
		\begin{tabular}{c r r c c}
			\midrule
			\textbf{Company} & \multicolumn{1}{c}{\textbf{$CAR$}} & \multicolumn{1}{c}{\textbf{$CAR$ Percentile}}  & \textbf{Impact} & \textbf{Event Period}\\
			\midrule
			& -0.015882584 & 26.9273 & None & [-1,0 ]\\
			& -0.027930361 & 19.52314 & None & [-1,1 ]\\
			MasterCard   & -0.025912927 & 27.83942 & None & [-1,3 ]\\
			& 0.107126511 & 97.48828 & Positive & [-1,5 ]\\
			& 0.135604103 & 97.36842 & Positive & [-1,10 ]\\
			\midrule
			& -0.028139803 & 12.72802 & None & [-1,0 ]\\
			& -0.040441177 & 9.26076 & Negative & [-1,1 ]\\
			Visa   & -0.047573044 & 11.8319 & None & [-1,3 ]\\
			& 0.140838043 & 99.09936 & Positive & [-1,5 ]\\
			& 0.112939429 & 94.97662 & Positive & [-1,10 ]\\
			\midrule
			& -0.024699949 & 19.20136 & None & [-1,0 ]\\
			& -0.024230469 & 24.77702 & None & [-1,1 ]\\
			Bank of America   & -0.031326586 & 24.99882 & None & [-1,3 ]\\
			& -0.092847266 & 3.46072 & Negative & [-1,5 ]\\
			& -0.128977999 & 2.344 & Negative & [-1,10 ]\\
			\midrule
			& 0.000824461 & 51.7293 & None & [-1,0 ]\\
			& 0.00794087 & 65.27714 & None & [-1,1 ]\\
			Vodafone   & 0.004882324 & 57.58806 & None & [-1,3 ]\\
			& -0.012009277 & 35.3259 & None & [-1,5 ]\\
			& -0.011377693 & 39.66936 & None & [-1,10 ]\\
			\midrule
			& -0.027029116 & 11.10264 & None & [-1,0 ]\\
			& -0.023728852 & 18.3817 & None & [-1,1 ]\\
			Apple   & -0.005504079 & 42.55964 & None & [-1,3 ]\\
			& -0.005196594 & 44.34158 & None & [-1,5 ]\\
			& 0.001828533 & 51.35376 & None & [-1,10 ]\\
			\midrule
			& 0.00547585 & 73.01768 & None & [-1,0 ]\\
			& 0.014332099 & 89.90238 & None & [-1,1 ]\\
			AT\&T & 0.024144556 & 94.53076 & Positive & [-1,3 ]\\
			& 0.014879076 & 80.10654 & None & [-1,5 ]\\
			& 0.027903421 & 88.6526 & None & [-1,10 ]\\
			\midrule
			& 0.002251211 & 57.74708 & None & [-1,0 ]\\
			& 0.002846928 & 58.18658 & None & [-1,1 ]\\
			Wells Fargo  & 0.006975269 & 64.68026 & None & [-1,3 ]\\
			& 0.010258418 & 67.66082 & None & [-1,5 ]\\
			& 0.006787068 & 60.10972 & None & [-1,10 ]\\
			\midrule
			& -0.007305775 & 30.4058 & None & [-1,0 ]\\
			& 0.013550503 & 77.18096 & None & [-1,1 ]\\
			JP Morgan Chase  & 0.031340549 & 90.29374 & Positive & [-1,3 ]\\
			& 0.053073337 & 96.23408 & Positive & [-1,5 ]\\
			& 0.09046202 & 98.84216 & Positive & [-1,10 ]\\
			\midrule
			& -0.001108209 & 43.57652 & None & [-1,0 ]\\
			& -0.009229913 & 14.02556 & None & [-1,1 ]\\
			TD Canada Trust   & -0.005118312 & 32.22566 & None & [-1,3 ]\\
			& -0.013999826 & 13.95322 & None & [-1,5 ]\\
			& 0.023281975 & 91.56936 & Positive & [-1,10 ]\\
			\midrule
			& -0.00041386 & 51.144 & None & [-1,0 ]\\
			& -0.003047091 & 43.91128 & None & [-1,1 ]\\
			American Express   & 0.006055721 & 63.35356 & None & [-1,3 ]\\
			& 0.025735512 & 86.15674 & None & [-1,5 ]\\
			& 0.050253429 & 94.5455 & Positive & [-1,10 ]\\
			\midrule
		\end{tabular}
	\end{table}
	
	\begin{table}
		\centering
		\caption* {TABLE \ref{Full Results} continued.}
		\begin{tabular}{c r r c c}
			\midrule
			\textbf{Company} & \multicolumn{1}{c}{\textbf{$CAR$}} & \multicolumn{1}{c}{\textbf{$CAR$ Percentile}}  & \textbf{Impact} & \textbf{Event Period}\\
			\midrule
			
			& -0.053961291 & 2.95846 & Negative & [-1,0 ]\\
			& -0.06072401 & 4.1384 & Negative & [-1,1 ]\\
			International Netherlands Group   & -0.020229264 & 34.1261 & None & [-1,3 ]\\
			& -0.029848597 & 30.17358 & None & [-1,5 ]\\
			& -0.081706673 & 12.26994 & None & [-1,10 ]\\
			\midrule
			& 0.016841037 & 74.30582 & None & [-1,0 ]\\
			& -0.002526331 & 46.45178 & None & [-1,1 ]\\
			LinkedIn  & -0.017375856 & 34.262 & None & [-1,3 ]\\
			& -0.026284125 & 30.81402 & None & [-1,5 ]\\
			& -0.045621561 & 26.63628 & None & [-1,10 ]\\
			\midrule	
			& -0.017162556 & 18.47456 & None & [-1,0 ]\\
			& -0.024200351 & 16.51522 & None & [-1,1 ]\\
			Microsoft   & -0.034090835 & 15.74556 & None & [-1,3 ]\\
			& -0.015471464 & 36.68258 & None & [-1,5 ]\\
			& 0.034940648 & 76.79054 & None & [-1,10 ]\\
			\midrule
			& 0.007847170 & 62.39716 & None & [-1,0 ]\\
			& -0.007792594 & 44.4221 & None & [-1,1 ]\\
			Royal Bank of Scotland   & -0.014817017 & 40.55122 & None & [-1,3 ]\\
			& 0.048606528 & 79.92352 & None & [-1,5 ]\\
			& 0.026742735 & 65.6532 & None & [-1,10 ]\\
			\midrule
			& 0.004977842 & 64.09092 & None & [-1,0 ]\\
			& 0.014513902 & 80.37984 & None & [-1,1 ]\\
			JP Morgan Chase  & 0.000234979 & 49.82888 & None & [-1,3 ]\\
			& -0.014250438 & 29.04242 & None & [-1,5 ]\\
			& -0.031128703 & 18.06696 & None & [-1,10 ]\\
			\midrule
			& -0.000184061 & 46.84568 & None & [-1,0 ]\\
			& 0.016315624 & 80.3436 & None & [-1,1 ]\\
			Bank of America   & 0.017620533 & 76.0839 & None & [-1,3 ]\\
			& 0.004226565 & 55.21462 & None & [-1,5 ]\\
			& 0.025992899 & 75.85016 & None & [-1,10 ]\\
			\midrule
			& -0.007029030 & 29.70018 & None & [-1,0 ]\\
			& -0.009490565 & 28.69322 & None & [-1,1 ]\\
			Facebook   & 0.024868377 & 73.9342 & None & [-1,3 ]\\
			& 0.047184864 & 86.43622 & None & [-1,5 ]\\
			& 0.092061897 & 95.16446 & Positive & [-1,10 ]\\
			\midrule
			& 0.001928050 & 54.8177 & None & [-1,0 ]\\
			& 0.001096061 & 52.14442 & None & [-1,1 ]\\
			Activision Blizzard  & -0.016714484 & 24.34886 & None & [-1,3 ]\\
			& -0.006333017 & 41.35782 & None & [-1,5 ]\\
			& -0.062767985 & 4.28474 & Negative & [-1,10 ]\\
			\midrule
			& -0.000274242 & 47.75408 & None & [-1,0 ]\\
			& -0.016656729 & 24.76632 & None & [-1,1 ]\\
			Danske Bank   & -0.014954993 & 31.98618 & None & [-1,3 ]\\
			& 0.008732713 & 58.03892 & None & [-1,5 ]\\
			& -0.007350568 & 44.4036 & None & [-1,10 ]\\
			\midrule
			& -0.004439445 & 35.08586 & None & [-1,0 ]\\
			& -0.018229923 & 15.0597 & None & [-1,1 ]\\
			Storebrand  & -0.063078035 & 0.4896 & Negative & [-1,3 ]\\
			& -0.061395155 & 1.39984 & Negative & [-1,5 ]\\
			& -0.049772166 & 7.76122 & Negative & [-1,10 ]\\
			\midrule
			& 0.000963002 & 49.70118 & None & [-1,0 ]\\
			& 0.003381149 & 54.0087 & None & [-1,1 ]\\
			Gjensidige Forsikr   & -0.010505422 & 37.04708 & None & [-1,3 ]\\
			& -0.040286066 & 15.8641 & None & [-1,5 ]\\
			& -0.028966577 & 29.2466 & None & [-1,10 ]\\
			\midrule
			& 0.002407147 & 60.15424 & None & [-1,0 ]\\
			& 0.004563586 & 62.29666 & None & [-1,1 ]\\
			Sony Corporation   & 0.001822970 & 58.2152 & None & [-1,3 ]\\
			& -0.021498560 & 37.79418 & None & [-1,5 ]\\
			& 0.014746326 & 63.87318 & None & [-1,10 ]\\
			\midrule
		\end{tabular}
	\end{table}
	
	\begin{table}
		\centering
		\caption* {TABLE \ref{Full Results} continued.}
		\begin{tabular}{c r r c c}
			\midrule
			\textbf{Company} & \multicolumn{1}{c}{\textbf{$CAR$}} & \multicolumn{1}{c}{\textbf{$CAR$ Percentile}}  & \textbf{Impact} & \textbf{Event Period}\\
			\midrule
			& 0.021753872 & 88.64916 & None & [-1,0 ]\\
			& 0.016990920 & 78.1646 & None & [-1,1 ]\\
			Amazon   & -0.036447006 & 11.17068 & None & [-1,3 ]\\
			& -0.043888322 & 11.2798 & None & [-1,5 ]\\
			& -0.038297272 & 21.30828 & None & [-1,10 ]\\
			\midrule
			& -0.000709556 & 48.81262 & None & [-1,0 ]\\
			& -0.007735669 & 39.30916 & None & [-1,1 ]\\
			Activision Blizzard   & 0.004492289 & 56.1743 & None & [-1,3 ]\\
			& -0.002983879 & 48.68772 & None & [-1,5 ]\\
			& 0.087616411 & 92.7746 & Positive & [-1,10 ]\\
			\midrule
			& -0.003403517 & 44.67244 & None & [-1,0 ]\\
			& -0.010883831 & 36.797 & None & [-1,1 ]\\
			Sony Corporation   & -0.000200726 & 50.39402 & None & [-1,3 ]\\
			& 0.026361475 & 68.49948 & None & [-1,5 ]\\
			& 0.026080234 & 64.358 & None & [-1,10 ]\\
			\midrule	
			& 0.014832688 & 88.2169 & None & [-1,0 ]\\
			& 0.025240246 & 94.41064 & Positive & [-1,1 ]\\
			Rackspace   & 0.043310538 & 97.7079 & Positive & [-1,3 ]\\
			& 0.040679243 & 94.7607 & Positive & [-1,5 ]\\
			& 0.040587064 & 89.53928 & None & [-1,10 ]\\
			\midrule
			& -0.018099374 & 21.19232 & None & [-1,0 ]\\
			& -0.012956852 & 32.89316 & None & [-1,1 ]\\
			Microsoft   & 0.019714934 & 71.0867 & None & [-1,3 ]\\
			& 0.028748607 & 74.6767 & None & [-1,5 ]\\
			& -0.021794832 & 37.10918 & None & [-1,10 ]\\
			\midrule	
			& 0.017723392 & 92.3236 & Positive & [-1,0 ]\\
			& 0.026038429 & 94.92672 & Positive & [-1,1 ]\\
			Sony Corporation   & 0.029414053 & 92.04924 & Positive & [-1,3 ]\\
			& 0.04572172 & 96.4814 & Positive & [-1,5 ]\\
			& 0.037697846 & 87.84194 & None & [-1,10 ]\\
			\midrule	
			& 0.003252883 & 55.78888 & None & [-1,0 ]\\
			& 0.006038200 & 58.02368 & None & [-1,1 ]\\
			Alibaba Group   & 0.027994135 & 73.14122 & None & [-1,3 ]\\
			& 0.028993768 & 71.19542 & None & [-1,5 ]\\
			& 0.064338831 & 82.87406 & None & [-1,10 ]\\
			\midrule
			& -0.010079453 & 26.95334 & None & [-1,0 ]\\
			& 0.002073589 & 55.11578 & None & [-1,1 ]\\
			Nordea Bank   & -0.030816091 & 12.2077 & None & [-1,3 ]\\
			& -0.061577132 & 2.58652 & Negative & [-1,5 ]\\
			& -0.174724652 & 0.0002 & Negative & [-1,10 ]\\
			\midrule
			& 0.012003977 & 73.88932 & None & [-1,0 ]\\
			& -0.007382446 & 39.45172 & None & [-1,1 ]\\
			Facebook   & 0.034738393 & 85.06418 & None & [-1,3 ]\\
			& 0.030894154 & 78.7436 & None & [-1,5 ]\\
			& 0.028141425 & 71.76988 & None & [-1,10 ]\\
			\midrule
			& -0.001261847 & 48.43654 & None & [-1,0 ]\\
			& -0.007662918 & 39.32008 & None & [-1,1 ]\\
			Amazon   & -0.014995234 & 34.24676 & None & [-1,3 ]\\
			& -0.003671321 & 48.09398 & None & [-1,5 ]\\
			& 0.002219653 & 53.17424 & None & [-1,10 ]\\
			\midrule
			& -0.026102614 & 10.94388 & None & [-1,0 ]\\
			& -0.044632962 & 5.67572 & Negative & [-1,1 ]\\
			EA Sports   & -0.052965749 & 7.55248 & Negative & [-1,3 ]\\
			& -0.020028623 & 30.5396 & None & [-1,5 ]\\
			& -0.034627666 & 26.39366 & None & [-1,10 ]\\
			\midrule
			& 0.009412519 & 73.84972 & None & [-1,0 ]\\
			& 0.037493964 & 94.77486 & Positive & [-1,1 ]\\
			Liberty Global   & 0.099809937 & 99.5675 & Positive & [-1,3 ]\\
			& 0.101846496 & 99.21738 & Positive & [-1,5 ]\\
			& 0.106292715 & 98.23724 & Positive & [-1,10 ]\\
			\midrule
			& 0.000637514 & 52.07072 & None & [-1,0 ]\\
			& -0.018705128 & 30.38584 & None & [-1,1 ]\\
			Overstock.com   & -0.023468622 & 31.49814 & None & [-1,3 ]\\
			& -0.003278425 & 49.49494 & None & [-1,5 ]\\
			& 0.010503096 & 58.08244 & None & [-1,10 ]\\
			\midrule
		\end{tabular}
	\end{table}
\end{document}